\documentclass[aps,showpacs,twocolumn]{revtex4}
\usepackage{amsmath,amsfonts,bm}
\usepackage[dvips]{graphicx}
\begin{document}
\title{Coherent macroscopic quantum tunneling in boson-fermion mixtures}
\author{D. Mozyrsky}
\author{I. Martin}
\author{E. Timmermans}
\affiliation{Theoretical Division, Los Alamos National Laboratory, Los Alamos, NM 87545}
\date{\today}
\begin{abstract}
We show that the cold atom systems of simultaneously trapped Bose-Einstein
condensates (BEC's) and quantum degenerate fermionic atoms provide promising
laboratories for the study of macroscopic quantum tunneling.  Our theoretical
studies reveal that the spatial extent of a small trapped BEC immersed in a
Fermi sea can tunnel and coherently oscillate between the values of the
separated and mixed configurations (the phases of the phase separation
transition of BEC-fermion systems).  We evaluate the period, amplitude and
dissipation rate for $^{23}$Na and $^{40}$K-atoms and we discuss the
experimental prospects for observing this phenomenon.
\end{abstract}
 \maketitle

The tunneling of a macroscopic (or collective) variable of a many-body
system through a classically forbidden region, macroscopic quantum
tunneling (MQT),
is a phenomenon of fundamental interest \cite{Leggett} and a recurring
theme in a
variety of fields ranging from nuclear (fission) and condensed matter
physics
(e.g. quantum magnets \cite{qm}, SQUIDs) to quantum optics (macroscopic
Schrodinger cat
states \cite{qo} and beyond-standard limit measurements).  Nevertheless,
stringent tests
under well-understood and controlled conditions remain an
experimental challenge.
Cold atom gases, arguably the cleanest and best understood mesoscopic
systems
which, furthermore, offer unprecedented control knobs such as the
ability to vary
the inter-particle interactions \cite{Fesh}, now provide an intriguing
candidate laboratory
for the study of MQT.

The first cold atom MQT proposals \cite{theory1} suggested observing the
collapse of
a trapped dilute gas Bose-Einstein condensate (BEC) of mutually
attracting bosons.
However, the experimental results \cite{Hulet, Wieman} were either too
sensitive to
particle number to distinguish MQT from classical collapse \cite{Hulet},
or the analysis was complicated by more complex dynamics (such as
'clumping') \cite{Wieman}.
Evidence of coherence (of the many-body system taking on a linear
superposition of
states that correspond to the macroscopic variable residing on either
side of
the barrier) is even more difficult to gather.  Such coherence would
be more
readily observable in the MQT between ${\it long}$-${\it lived}$ states, in
which case one
could set up a coherent population oscillation between the many-body
states.
Such long-lived states naturally occur in (zero-temperature) first
order phase
transitions in which the order parameter, which provides the
macroscopic variable,
can tunnel through the barrier of its Landau-Ginzburg potential.  In
the infinite
system limit, the coupling between the two states rigorously
vanishes, but finite-
size cold atom systems of moderate particle numbers provide, once
again, a
promising candidate to observe the MQT coherence between states of
different
phases, as we show below.

An earlier proposal to observe MQT between states in which the
components of a
BEC-mixture arrange themselves differently in space, involved a very
low coupling
on account of the small spatial overlap between the single component
densities
in the different states \cite{Kasamatsu}.  In this paper, we propose that
MQT can be
realized and its coherence, perhaps, observed in trapped gas mixtures of
a single-component fermion system and a BEC.  Such mixtures are
currently created
\cite{sympath} e.g. in the sympathetic cooling scheme in which the colder
BEC cools the
fermions.  The tunneling and coherent oscillations that we target
would occur between
states of the mixed and separated phases in the phase separation
transition
of the fermion-BEC mixture \cite{Molmer}.  Such transitions could
be accessed
by varying the scattering length of the boson-fermion interaction
\cite{Simoni}.

We consider $N_B$
atomic bosons confined in a spherically symmetric harmonic trap
(of frequency $\omega_T$) interacting with a much larger system of atomic
fermions. For simplicity we assume the fermions to occupy
an infinite volume.
The Hamiltonian of the bosons is described by the standard
Gross-Pitaevskii (GP) form \cite{Leggett}, i.e., with inter-particle
interactions described by a contact potential ($\propto \lambda_{BB} \delta({\bf r}-{\bf r}')$),
which we choose to be repulsive ($\lambda_{BB}>0$)
We assume that the interaction of bosons with fermions is also contact-like, contributing
$\lambda_{BF}|\Psi_B|^2|\Psi_F|^2$ to the Hamiltonian density,
where $\lambda_{BF}$ is the fermion-boson coupling constant.
Furthermore, all fermions occupy in the same spin state
so that the short-range inter-fermion interactions do not contribute by virtue of the
Pauli exclusion principle.

We are interested in the dynamics of the reduced system of bosons described by the functional
\begin{eqnarray}
S=S_{BEC}+{\rm Tr}\log\left[\hbar\partial_\tau-{\hbar^2\nabla^2\over
2m_F}-\mu_F+\lambda_{BF}|\Psi_B|^2\right],\nonumber\\\label{2}
\end{eqnarray}
where $S_0$ is the action of the bosons alone, $S_{BEC}=\int d\tau(\hbar\int d^3{\bf
r}{\dot\Psi}_B\Psi_B^\ast - H_{BEC})$, and the second term
is a contribution due to the interaction of bosons with fermions;
$\mu_F$ is the chemical potential of the fermions.
Here and throughout the paper we will be utilizing the imaginary time (Matsubara)
representation, unless stated otherwise. An explicit evaluation of the second term
is a challenging task. However, here we are interested in the dynamics of the slow
breathing mode of the BEC $\Psi_B^0$, which can be treated in the self-similar
density approximation. This dynamics describes the
longitudinal expansions (and contractions) of the condensate.  Finite size
effects such as the appearance of a non-vanishing excitation energy (gap)
can decouple this mode from other excitation modes. Hence, $\Psi_B^0$ peaks at
small frequencies ($\omega$) and small wavevectors (${\bf q}$), giving a
$\Psi_B^0$ that is a slowly varying function of spacial and temporal
coordinates. In such case the ${\rm Tr}\log[...]$ in
Eq.~(\ref{2}) can be evaluated within the Thomas-Fermi approximation.
A straightforward zero-temperature calculation yields
\begin{eqnarray}
{\delta{\rm Tr}\log[...]\over\delta\Psi_B^\ast} =
{\lambda_{BF}k_F^3\over 3\pi^2}\,{\rm Re}\left[1 -
{\lambda_{BF}|\Psi_B({\bf r})|^2\over
\mu_F}\right]^{3/2}\Psi_B,\label{3}
\end{eqnarray}
where $k_F$ is the Fermi wavevector. Eq.~(\ref{3}) represents an
additional term in the Gross-Pitaevskii (GP) equation, $\delta
S_{GP}/\delta\Psi_B^\ast=0$, resulting from interaction with
fermions. In order to analyze the physical meaning of Eq.~(\ref{3})
let us expand it in powers of $\Psi_B$. The first nontrivial
contribution is a term $-2\lambda^\prime|\Psi_B|^2\Psi_B$, $\lambda^\prime=(\lambda_{BF}^2
k_F^3/4\pi^2\mu_F)$, which corresponds to the
attraction between bosons mediated by interaction with fermions.
For nonzero, but small $\omega$ and $\bf q$ there is an additional term
(of the order of $\lambda_{BF}^2$) related to the dissipation of the
condensate due to the Landau damping, as we discuss below. The next order yields
$\eta|\Psi_B|^4\Psi_B$, $\eta=(k_F^3\lambda_{BF}^3/8\pi^2\mu_F^2)$.  Unlike
the previous term this one is positive, and represents reduction
in the effective boson-boson attraction due to depletion of
fermions in the regions of high density of the bosons. The next
order terms (in $\lambda_{BF}$) prove to be unimportant as can
be verified directly from Eq.~(\ref{3}). Therefore we will replace the potential energy contribution in
GP equation given by Eq.~(\ref{3}) by the two terms discussed
above \cite{com}.

To analyze the dynamics of the slow (breathing) mode
described by the Hamiltonian
\begin{eqnarray}
H = \int d^3{\bf r}\Big[{\hbar^2\over 2m_B}|\nabla\Psi_B|^2
+{m_B\omega_T^2 {\bf r}^2\over 2 }|\Psi_B|^2~~~~~~~~~~\label{4}\\
+{1\over 2}(\lambda_{BB}-\lambda^\prime)|\Psi_B|^4+{\eta\over 6}
|\Psi_B|^6\Big],\nonumber
\end{eqnarray}
we apply the time dependent variational principle. Since we are interested in
ground state properties of Eq.~(5) we use a spherically
symmetric Gaussian trial wavefunction
\begin{eqnarray}
\Psi_B^0({\bf r}) = {N_B^{1\over 2}\over \pi^{3\over
4}(xR_0)^{3\over 2}}\exp\left[-{{\bf r}^2\over
2(xR_0)^2}\right],\label{6}
\end{eqnarray}
parameterized by a dimensionless parameter $x$ that
characterizes the BEC's spatial width in units of the zero point motion amplitude
$R_0 = (\hbar/2m_B\omega_T)^{1/2}$. Substitution of this
wavefunction into Eq.~(5) yields the following dependence of the
ground-state energy $E^0$ on $x$:
\begin{eqnarray}
E_0(x) = {3N_B\hbar\omega_T\over 2}\left({1\over x^2} +{x^2\over 4}-{\alpha\over 3x^3} +{\beta\over 6x^6}\right),\label{7}
\end{eqnarray}
where $\alpha =
N_B(\lambda^\prime-\lambda_{BB})/[(2\pi)^{3/2}R_0^3\hbar\omega_T]$
and $\beta = 4N_B^2\eta/(3^{5/2}\pi^3 R_0^6\hbar\omega_T)$.
\begin{figure}[h]
\vspace{-0 mm}
{\includegraphics[width = 3  in, angle=0]{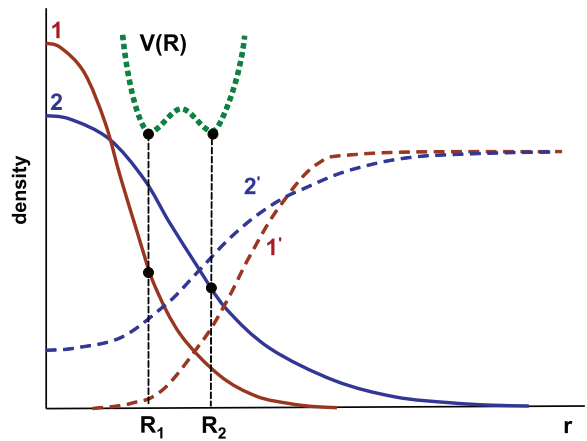}}
\caption{Density profiles of the bosons (solid lines 1 and 2) and corresponding
density profiles of the fermions (dashed lines 1' and 2') in two metastable states:
(1) with fermions having zero density at the center of the trap (``separated phase'') and
(2) with nonzero density of fermions (``mixed phase''). The dotted line represents schematically
an effective potential for the breathing mode of the bosons.} \vspace{-0 mm}
\end{figure}
For positive but
relatively small $\alpha$, i.e., for $\alpha
<\alpha_{cr}=32(2/5)^{1/4}/15 \simeq 1.69$, $E_0$ may develop two
competing minima, depending on the value of the $\beta$-parameter.
The energy barrier separating the minima is caused by the same
effect as the barrier appearing in the description of a
BEC with attractive interactions: it arises due to the competition
between the kinetic and the interaction energies, i.e, the first
and the third terms in the right-hand side (rhs) of Eq.~(\ref{7}).
In the absence of the last term in the rhs of Eq.~(\ref{7}) the
state in this well would have been metastable - the energy would tend
to $-\infty$ at $x\rightarrow 0$. The $1/x^6$ term
stabilizes the system: for small $x$ this term rapidly
increases, giving rise to another minimum of $E_0(x)$, now due to the competition
between the last two terms in the rhs of Eq.~(\ref{7}).

At certain values of $\alpha$ and $\beta$ the two minima of $E_0(x)$ will have the same energy, and
the ground state
of the system becomes degenerate. Since our system is finite, this degeneracy will be lifted
by the quantum tunneling transition between the two states. Such mechanism has been
suggested to be the dominant decay process for condensates with attractive interactions between particles \cite{theory1}.
The tunneling corresponds to the low energy excitations of the breathing mode, described by the wavefunction in Eq.~(\ref{6}).
It has been shown in \cite{theory1} that by accounting for the superfluid motion of the condensate
(which can be done by
introducing a phase-factor $e^{i\phi}$ for the wavefunction in Eq.~(\ref{6}) and requiring the superfluid velocity ${\bf v}_s = (\hbar/m_B){\bf\nabla}\phi$
to satisfy the continuity equation) one obtains an effective action for the breathing mode
of the condensate
\begin{eqnarray}
S_0[x(\tau)]=\int d\tau\left[{m_0 {\dot x}^2\over 2}+E_0(x)\right],\label{10}
\end{eqnarray}
where $E_0(x)$ is given by Eq.~(\ref{7}) and $m_0 = 3m_B N_B
R_0^2/2$. Thus the dynamics of the ground state wavefunction of
the condensate is that of a quantum particle of mass $m_0$ moving in the
potential $E_0(x)$.

A direct analysis of the Shr{o}dinger equation corresponding to Eq.~(\ref{10}), however, is quite
cumbersome since the two wells
are generally quite asymmetric. Instead we choose an alterative
route: we compute the ground state energy and obtain the tunneling rate
by numerically solving the time-independent GP equation, $\delta
H/\delta\Psi_B=E\Psi_B$, where $H$ is given by Eq.~(\ref{4}). The latter
approach also serves as an independent justification of the
variational method and confirms that macroscopic quantum tunneling, QMT, is
the mechanism that causes the transition between the two states of the
condensate. Upon substitution $\Psi_B \sim\phi/r$, the
time-independent GP equation can be cast in the form
\begin{eqnarray}
\left[-{\partial^2\over\partial x^2}+ {x^2\over 4}-a{\phi^2\over
x^2}+{b\over 4}{\phi^4\over x^4} \right]\phi=\mu\phi,\label{11}
\end{eqnarray}
where the $\phi(x)$-function is normalized to unity,
$a=(\pi/2)^{1/2}\alpha$, $b=3^{5/2}\pi\beta/16$, and $x=r/R_0$, $\mu=E/\hbar\omega_T$.
We find the ground state numerically by replacing the rhs of
Eq.~(\ref{11}) by $-\partial_\tau\phi$ and propagating $\phi$ in
the imaginary time $\tau$ until it converges to the ground state
$\phi_0$ (or $\Psi_B^0$). We then evaluate the ground state energy
according to Eq.~(\ref{4}) and present the results in Fig.
2(a) as a function of the $b$-parameter for different values of
$a$. Fig. 2(b) shows the dispersion of the ground state width,
$(1/N_B)\int d^3{\bf r} |\Psi_B^0({\bf r})|^2{\bf r}^2$, as
a function of those same parameters. For $a<a_{cr}=1.83$ the ground
state energy and dispersion undergo a sharp crossover between the
state with compressed and expanded BEC wavefunctions
(corresponding to the phase separated and mixed states) as
functions of $b$. Note that the value $a_{cr}$ corresponds to the
value of $\alpha_{cr}=1.46$, which is quite close to the above
critical value of $1.69$ obtained from the variational
approach. The dependence of ground state energy near the critical
value of $a$ is shown in the inset of Fig. 2(a). Clearly
the ground state energy exhibits {\it avoided level crossing}, which is in accordance with
the above conjecture (e.g., Eq.~(\ref{10})) of macroscopic quantum tunneling
between the two local energy minima.
\begin{figure}[h]
{\includegraphics[width = 3.3  in, angle=0]{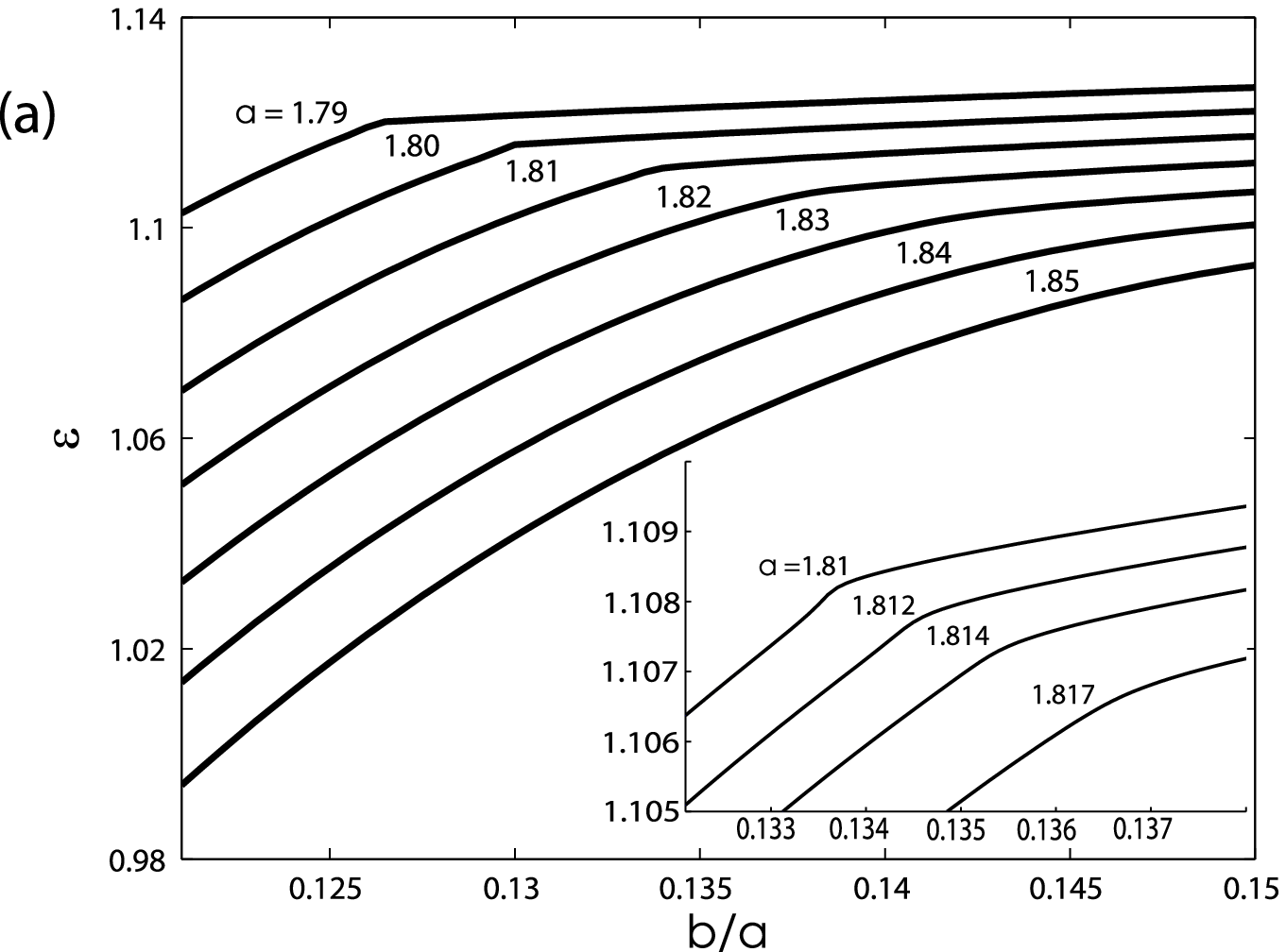}}

\vspace{5 mm}
{\includegraphics[width = 3.3  in, angle=0]{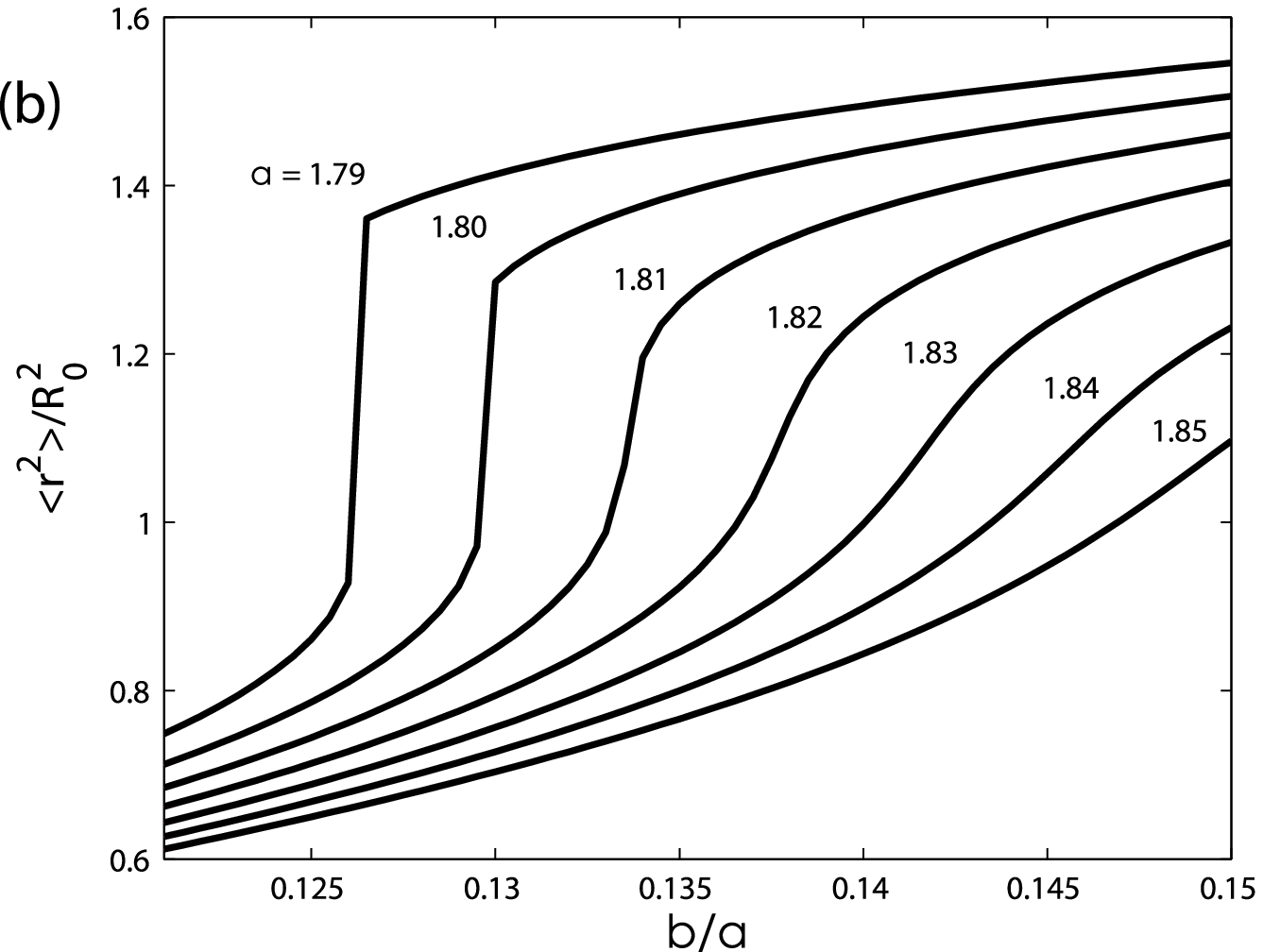}}
\caption{(a) Dependence of the ground state energy of the BEC (per particle, in units of $\hbar\omega_T$)
as a function of parameters $a$ and $b$; (b) Dispersion of the ground state
spatial extent as a function of the same parameters.} \vspace{-0 mm}
\end{figure}

The value of the tunneling matrix element $\Delta$ between two local ``ground'' states $\epsilon_1$
and $\epsilon_2$ can be deduced by fitting the calculated energy curves in Fig. 1 with the standard expression \cite{Landau},
$\epsilon = (\epsilon_1+\epsilon_2)/2-[(\epsilon_1-\epsilon_2)^2/4-\Delta^2]^{1/2}$
and assuming that in the vicinity of the point of crossover both $\epsilon_1$ and $\epsilon_2$ are linear
functions of parameter $b$. For $a=1.81$ one finds $\Delta\sim 10^{-4}\times \hbar\omega_T$, while for
$a=1.82$, $\Delta\sim 10^{-2}\times \hbar\omega_T$. Assuming that the ground state wavefunctions
have Gaussian shape, $|\Psi_B|^2_{1(2)}\sim R^{-3}_{1(2)}\exp(-{\bf r}^2/R^2_{1(2)})$, from Fig. 2 one finds
that ${\bar R}=(R_1+R_2)/2\simeq 0.85 R_0$ for both $a=1.81$ and $a=1.82$, and $\delta R=|R_1-R_2|\simeq 0.09 R_0$
for $a=1.81$ and $\delta R \simeq 0.03 R_0$ for $a=1.82$. For a typical value of the trapping frequency $\nu_T = 10^2 Hz$
($\omega_T=2\pi\nu_T$), the two tunneling rates are $\Delta_{1.81}/\hbar=10^{-2}\,s^{-1}$ and $\Delta_{1.82}/\hbar=10^{2}\,s^{-1}$.
Since the value of $R_0$ for most trapped atomic BEC's is of the order of a few microns, the difference between the
radii of the two condensate states $\delta R$ is submicron.
Such small variation may be difficult to observe in situ by optical means.  However, the
expansion process that takes place in time-of-flight measurements after the trap potential
is shut off and the expanding atoms are observed, has successfully magnified small distance features
in other experiments.

{\it Role of dissipation:} The above analysis determines the tunneling rate, but does not address the
question whether the tunneling process is quantum coherent.  Will the probability of the system
to occupy one of the two macroscopic states oscillate in time as $\cos^2{(\Delta\, t/\hbar)}$? The fermions not only provide the BEC with the effective interaction, they also cause fluctuations
which can destroy the macroscopic quantum coherence. To evaluate the effect of fluctuations,
it is sufficient to consider the first non-vanishing frequency-dependent
contribution into the effective action of the bosons coming from the perturbative expansion of the
${\rm Tr\log{[...]}}$ term in
Eq.~(\ref{2}):
\begin{eqnarray}
-{\lambda_{BF}^2\over 2\hbar}\int {d\omega\over 2\pi}\int {d^3{\bf q}\over (2\pi)^3}\chi_0({\bf q},\omega)|\rho_B(\omega,{\bf q})|^2.\label{12}
\end{eqnarray}
Here $\rho_B({\bf q},\omega)$ is the Fourier transform of $\rho_B({\bf r},t)$ and $\chi_0$ is the response function of the non-interacting
fermions. In the small frequency domain $\chi_0 = (1/4\pi)[\hbar^2 k_F^3/(\pi\mu_F)+m_F^2|\omega|/(\hbar^2 q)]$. The frequency-independent
part of $\chi_0$ has already been incorporated in the effective interaction between bosons, i.e.,
$\lambda^\prime |\Psi_B|^4$ term in  Eq.~(\ref{4}).
The second term in $\chi_0$ is responsible for damping. To quantify its role we employ a
two-state approximation in describing the tunneling dynamics.  In this representation
the tunneling is described by the Hamiltonian $H_{\rm tun} = \Delta{\hat\sigma}_x$,
where ${\hat\sigma}_x$ is a Pauli matrix with non-zero off-diagonal
elements, and the position operator, i.e. the spatial width of the ground-state BEC wavefunction,
is given by ${\hat R}={\bar R}+(\delta R/2){\hat\sigma}_z$,
where ${\hat\sigma}_z$ is the diagonal Pauli matrix (with $\pm 1$ along the diagonal). The dissipative part of the action for $H_{\rm tun}$ can be
derived from Eq.~(\ref{12}) by substituting a Gaussian ansatz,
$\rho_B({\bf r},t)=N_B/[\pi^{3/2}R^3(t)]\exp{[-{\bf r}^2/R^2(t)]}$, where $R(t)= {\bar R}+(\delta R/2)\sigma_z(t)$, $\sigma_z=\pm 1$, into Eq.~(\ref{12}).
For $\delta R \ll {\bar R}$ one obtains
\begin{eqnarray}
S_{\rm diss}=\gamma\hbar\int d\tau d\tau^\prime \sigma_z(\tau)\sigma_z(\tau^\prime) (\tau-\tau^\prime)^{-2},\label{13}
\end{eqnarray}
where $\gamma = N_B^2\lambda_{BF}^2 m_F^2\delta R^2/[2(2\pi\hbar)^4 {\bar R}^4]$. Eq.~(\ref{13}), together with $H_{\rm tun}$ defined above, describes dissipative dynamics of a
two-state system. Such dynamics has been extensively studied in connection with macroscopic quantum tunneling of a superconducting phase in Josephson junctions,
and is known to depend critically on the value the parameter $\gamma$.
Specifically, for  $\gamma >1$ the two-state oscillation is always overdamped
and at zero temperature it exhibits localization as a result of quantum fluctuations \cite{leggett}.
It is therefore instructive to
evaluate $\gamma$ for our situation. For estimates we consider an atomic mixture of $^{23}$Na (bosons) and $^{40}$K (fermions), which have natural scattering lengthes
$a_{BB}\simeq 1\,nm$ ($\lambda_{BB}=4\pi\hbar^2 a_{BB}/m_B$) and $a_{BF}\simeq 4\,nm$ ($\lambda_{BF}=2\pi\hbar^2 a_{BF}[(1/m_B)+(1/m_F)]$). For these data we obtain a critical value of
$N_B^{cr}\simeq 12400$ (again for $\nu_T = 10^2 Hz$) and the fermion density $n_F^{cr}\simeq 7.4\times 10^{15} cm^{-3}$. Then, for $a=1.81$ we obtain $\gamma_{1.81}\simeq 1.1$, which
corresponds to the localized case (at $T=0$), whereas for $a=1.82$ one gets $\gamma_{1.82}\simeq 0.1$. In the high temperature limit (for $k_B T>\Delta$) the relaxation rate $\Gamma$ can
be expressed in terms of $\gamma$ as $\hbar\Gamma = \pi\gamma k_B T$ \cite{leggett},
and therefore coherent (underdamped) oscillations can be observed for
$T\ll \Delta_{1.82}/(\gamma_{1.82} k_B) = 0.5 nK$. The situation can be improved, however, if one utilizes a Feshbach resonance \cite{Fesh} to increase the $a_{BF}$ scattering length. For example,
for $a_{BF}=80nm$ one finds $N_B^{cr}\simeq 25$ and $n_F^{cr}\simeq 2.6\times 10^{11} cm^{-3}$, and $\gamma_{1.82}\simeq 2.5\times 10^{-4}$. For such parameters coherent oscillations
can be observed for $T\ll 0.2 \mu K$, which is easily observable. A low particle number also
reduces the uncertainty of an atomic counting measurement that can be carried out in the time-of-flight
procedure \cite{25}.

In summary we argue that a trapped boson-fermion mixture can exhibit MQT tunneling and
coherent oscillations. Our studies indicate that MQT can be observed in $^{23}$Na and
$^{40}$K atomic mixtures of sufficiently low temperatures.

We thank M. Boshier and S. A. Gurvitz for valuable discussions. The work is supported
by the US DOE.


\begin{thebibliography}{99}
\bibitem{com} A quantitaive analysis of the GP equation for
the potential given by Eq.~(2) for will be presented elsewhere.
\end{thebibliography}
\end{document}